\documentclass[prb,twocolumn,showpacs,superscriptaddress,preprintnumbers,amssymb]{revtex4}
\usepackage{graphicx}
\usepackage{dcolumn}
\usepackage{bm}
\usepackage{wasysym}

\newcommand{\beq}{\begin{equation}}
\newcommand{\eeq}{\end{equation}}
\newcommand{\beqn}{\begin{eqnarray}}
\newcommand{\eeqn}{\end{eqnarray}}

\begin{document}
\title{Fractionalization in Josephson Junction Arrays Hinged by Quantum Spin Hall edges}

\author{Cenke Xu}
\author{Liang Fu}
\affiliation{Department of Physics, Harvard University, Cambridge,
MA 02138}
\date{\today}

\begin{abstract}

We study a novel superconductor-ferromagnet-superconductor
(SC-FM-SC) Josephson array deposited on top of a two-dimensional
quantum spin Hall insulator. The Majorana bound state at the
interface between SC and FM leads to charge-$e$ tunnelling between
neighboring superconductor islands, in addition to the usual
charge-$2e$ Cooper pair tunnelling. Moreover, because Majorana
fermions encode the information of charge number parity, an exact
$Z_2$ gauge structure naturally emerges and leads to many new
phases, including a deconfined phase where electrons fractionalize
into charge-$e$ bosons and topological defects. A new SC-insulator
transition has also been found.

\end{abstract}
 \maketitle

\section{Introduction}

Superconductor Josephson junction arrays have been studied
extensively in the past. At low temperature, the phase of a
superconductor island becomes a quantum degree of freedom, which
is conjugate to the Cooper pair density. Competition between
charging energy and Josephson tunnelling leads to a
superconductor-insulator transition at zero temperature, which is
usually studied using quantum rotor models. Such a boson-only
approach is no longer adequate if low-energy quasi-particles are
present. Recently, it was proposed that if a $s$-wave
superconductor (SC) and ferromagnet (FM) junction is hinged by the
edge of a quantum spin Hall (QSH) insulator discovered recently
\cite{Konig2007}, a zero-energy Majorana bound state will be
localized at the SC-FM interface \cite{fukane}. Although the
Majorana fermions do not carry the global U(1) charge of
electrons, they do encode the information of fermion parity $i.e.$
the even-odd of the electron number, hence Majorana fermion will
participate in the tunnelling of charges between two SC islands,
and enrich the physics of the Josephson array.

So far most work on topological insulators have been focusing on
the band structure or weak interaction effects. Recently strong
correlation effects for topological insulators have attracted more
and more attentions \cite{balents2009,xu2010a,xu2010b}. Although
it has been proposed that quantum computation may be realized
based on Majorana bound states localized in topological defects
\cite{quantumcomputing}, very little was studied about the
many-body or strong correlation effects that the Majorana fermions
can participate in. In this work we will focus on the correlation
physics in the Majorana fermion assisted Josephson array, and we
will show that such Josephson arrays have unusual fractionalized
phases and phase transitions.

\section{one dimensional case}

\begin{figure}
\includegraphics[width=2.7in]{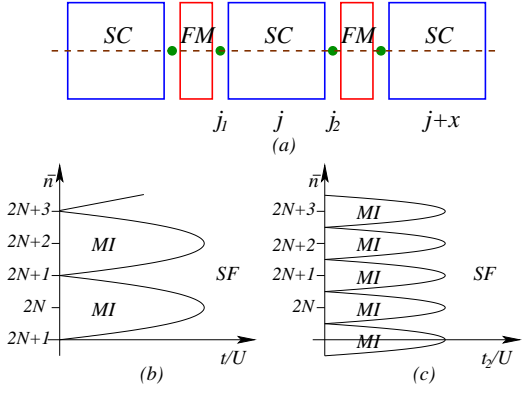}
\caption{($a$), the lattice structure for case 1. The green
circles are the Majorana fermions, and the dashed brown line is
the edge of the underlying QSH insulator. ($b$), the phase diagram
for ordinary SC Josephson array. ($c$), the phase diagram for
Josephson array in our case, where the number of MI lobes doubles
compared with ($b$). } \label{scm}
\end{figure}

Let us warm up with a simple one dimensional geometry, as depicted
in Fig.~\ref{scm}$a$. We denote the location of every SC island by
coordinate $j$, and denote the two Majorana fermions around each
SC island as $\gamma_{j,1}$ and $\gamma_{j,2}$. The FM islands
have uniform magnetizations that are perpendicular to the angular
momentum carried by the electrons of the QSH edge states, which
opens up a gap for the QSH edge states. As was pointed out by
Ref.~\cite{fu2009b}, these two Majorana fermions correlate with
the fermion number on this SC island through the constraint $
i\gamma_{j,1}\gamma_{j,2} = (-1)^{n_j}$. The tunnelling between SC
islands has to be consistent with this constraint, and the
following two terms are allowed \cite{andreev}: \beqn H_{t1} &=&
\sum_{j} - t_1 \cos(\phi_j - \phi_{j + x}), \cr\cr H_{t2} &=&
\sum_{j} - t_2 i \gamma_{j,2}\gamma_{j + x,1}\cos[(\phi_j -
\phi_{j + x})/2]. \eeqn Here $\phi_j$ is the phase angle of the SC
island $j$, and $e^{i \phi_j}$ increases the electron number $n_j$
by 2. $H_{t1}$ is the ordinary Josephson tunnelling term, and
$H_{t2}$ is the charge-$e$ tunnelling assisted by the Majorana
fermions, which is now allowed because with the Majorana fermions
there is no longer a Cooper pair breaking gap between even and odd
electron number on each SC island \cite{fu2009b} $i.e.$ electron
can reside across the SC island nonlocally through majorana zero
modes $\gamma_{j,1}$ and $\gamma_{j,2}$. Since $H_{t1}$ is a
second order effect that involves a Cooper pair breaking
intermediate state, in the limit with dominant Cooper pair energy,
$H_{t1}$ is ignorable. Inclusion of small $H_{t1}$ will only
quantitatively change the physics discussed in this paper.

It is well-known that the one dimensional Majorana fermion is
equivalent to a transverse field quantum Ising model, and the
Ising variables are defined on the links of the 1d lattice denoted
as $(j, j + x)$ in Fig.~\ref{scm}: \beqn \sigma^x_{j, j + x} =
\prod_{k\leq j} i\gamma_{k,1}\gamma_{k, 2}, \ \ \sigma^z_{j, j +
x} = i\gamma_{j,2}\gamma_{j + x,1}. \eeqn Now the full Hamiltonian
can be written as \beqn H = \sum_j U(n_j - \bar{n})^2 - t_2
\sigma^z_{j,j+x} \cos(\frac{\phi_j}{2} - \frac{\phi_{j + x}}{2}),
\label{1dgauge}\eeqn which is subject to the constraint \beqn
\sigma^x_{j-x, j}\sigma^x_{j, j+x}(-1)^{n_j} = 1. \eeqn The $U$
term represents a charging energy. The Hamiltonian Eq.
\ref{1dgauge} takes the standard form of a $Z_2$ gauge field
$\sigma^z$ coupled to matter field $\phi/2$. For this 1D system
the $Z_2$ gauge field $\sigma^z_{j,j+x}$ can be absorbed into the
rotor variable $\phi_j$ through the following duality mapping:
\beqn \sigma^z_{j,j+x} = \tau^z_j \tau^z_{j+x}, \ \theta_j =
\frac{\phi_j}{2} + \pi \frac{1 - \tau^z_j}{2}, \
\sigma^x_{j,j-x}\sigma^x_{j,j+x} = \tau^x_j. \label{1ddual}\eeqn
$\tau^z_j, \tau^x_j = \pm 1$ are Ising operators defined on SC
islands, and they satisfy the algebra of Pauli matrices.
$\theta_j$ and $n_j$ satisfy the rotor phase-number algebra:
$[\theta_j, n_k] = i\delta_{jk}$. Now this model can be written
with the new variables as ordinary Bose-Hubbard model
\cite{bosehubbard} in 1D: \beqn H = \sum_j U(n_j - \bar{n})^2 -
t_2\cos(\theta_j - \theta_{j+1}), \label{BHmodel}\eeqn and the
gauge constraint operator
$\sigma^x_{j,j-x}\sigma^x_{j,j+x}(-1)^n_j = \tau^x_j(-1)^{n_j}$
commutes with $e^{i\theta_j}$.

The phase diagram of the model Eq. \ref{1dgauge} is identical to
model Eq.~\ref{BHmodel}: with integer $\bar{n}$, there is a
Mott-insulator (MI) phase with fixed fermion number on every SC
island when $t_2/U \ll 1$, and a superfluid (SF) phase when $t_2/U
\gg 1$. The phase diagram is shown in Fig.~\ref{scm}$c$. When
$\bar{n}$ is precisely half-integer, since in this model there is
no nearest neighbor density repulsion, the MI phase vanishes.
Compared with the ordinary Josephson array, the number of MI lobes
is doubled in our case (Fig.~\ref{scm}$b$, $c$), basically because
there is no charge gap between even and odd particle filling. The
MI phase is nondegenerate, every SC island has fixed charge
number. In the SF phase, besides the Cooper pair operator
$e^{i\phi}$, the following string operator also has algebraic
correlation: \beqn \langle e^{ i\phi_j/2}\prod_{m = 1}^{L}
\sigma^z_{j, j + m} e^{- i \phi_{j + L}/2} \rangle \sim \langle
e^{i\theta_j}e^{-i\theta_{j+L}} \rangle \sim (1/L)^K. \eeqn with
the Luttinger parameter $K$ tuned by $t_2/U$.

Single electron can be injected into the system through the
Majorana fermion bound state. One legitimate representation of
electron operator is $c_{j,1} \sim e^{i\phi_j/2}\gamma_{j,1}$
\cite{andreev,fu2009b}. The spin index does not appear in the
electron operator because the spin degeneracy is lifted by FM
islands and the spin-orbital couplings in underlying QSH edge
states. Also, under transformation $\phi_j \rightarrow \phi_j +
2\pi$, the solution of the Majorana bound state $\gamma_{j,a}$
changes sign,
therefore the physical electron operator remains invariant. In
terms of the bosonic variables, the electron operator can be
expressed as \beqn c_{j, 1} \sim \prod_{k \leq j} \sigma^z_{k - x,
k }\sigma^x_{j - x, j} e^{i\phi_j/2} \sim \exp[i\pi\sum_{k < j}
n_k] e^{i \theta_j}. \label{electron}\eeqn The $\sigma^x$ in the
product Eq. \ref{electron} guarantees the fermionic statistics
between two electron operators. The correlation function between
two electron operators is \beqn \langle c_{j-L,
1}c^\dagger_{j,1}\rangle \sim \langle e^{i\theta_{j-L}} \exp[i\pi
\sum_{k = j-L}^{L - 1} n_k] e^{i\theta_j}\rangle \sim
(\frac{1}{L})^{K + \frac{1}{4K}}.\eeqn The Bosonic representation
of electron operator Eq. \ref{electron} takes exactly the same
form as the standard fermionization of the Bose variables $\theta$
and $n$ in one dimension.

The phase transition at the integer filling is a
Kosterlitz-Thouless (KT) transition at $K = 1/4$, which physically
corresponds to proliferating $2\pi$ kinks of $\theta$ defined in
Eq. \ref{1ddual}, which is equivalent to a $4\pi$ kink of $\phi$.
At this transition, the scaling dimension of Cooper pair operator
is $1/2$, while in ordinary 1D Josephson array, at the KT
transition the Cooper pair operator has scaling dimension 1/8,
because normally the transition is driven by the proliferation of
$2\pi$ kink of $\phi$. The difference between these two cases
stems from the charge-$e$ tunnelling enabled by the Majorana
fermions. In terms of $\theta$ introduced in Eq.~\ref{1ddual},
$H_{t1}$ is simply $\sum_{j} - t_1 \cos(2\theta_j - 2\theta_{j +
x})$, therefore turning on $H_{t1}$ in Eq.~\ref{BHmodel} will not
change the phase diagram, as long as there is no pairing gap
between even and odd filling. The random one-dimensional lattice
of Majorana fermions at the edge of a topological insulator, which
is similarly connected to the random Ising model, has been
discussed \cite{moorechain3,moorechain4}.


\section{two dimension, regular structure}

\begin{figure}
\includegraphics[width=3.0in]{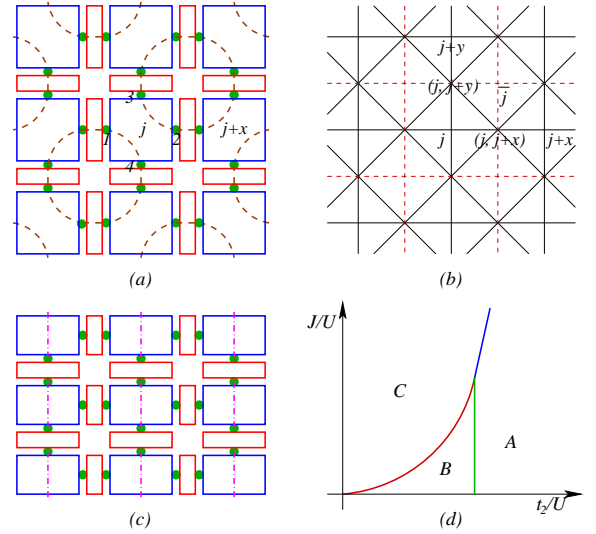}
\caption{$(a)$, the Josephoson lattice structure of case 2. $(b)$,
the effective lattice of the Josephson array in $(a)$. The SC
islands are denoted by $j$, and the $Z_2$ gauge field $\sigma^x$
and $\sigma^z$ are defined on links $(j, j+\nu)$. The dual Ising
variables $\tau^a$ are defined on the sites $\bar{j}$ of the dual
lattice, represented by red dashed lines. $(c)$, case 3 with
compressed $y$ direction, the intra-island tunnelling
$Ji\gamma_{j,3}\gamma_{j,4}$ is denoted by the pink dashed lines.
$(d)$, the phase diagram of case 3 with integer $\bar{n}$, plotted
against $t_2/U$ and $J/U$. Phase $A$, $B$ and $C$ represent the
SF, the $Z_2$ liquid, and the $\sigma^x$ ordered phase
respectively. The green line between phase $A$ and $B$ is a $3d$
XY transition, and the transition between $B$ and $C$ is a first
order transition described by Eq. \ref{xumoore}. The direct
transition between phase $A$ and $C$ may expand into a stable
roton liquid phase.} \label{scm2}
\end{figure}

Now let us move on to the 2D structure as depicted in
Fig.~\ref{scm2}$a$. We place the SC and FM islands on top of a QSH
insulator with commensurate holes. The internal edges of these
holes hinge all the islands in this lattice. We first assume that
the FM island is very thin compared with SC island, therefore the
intra-island tunnelling between Majorana fermions is negligible
compared with inter-island tunnellings. We denote every SC island
as site $j$, and denote the four Majorana fermions around each SC
island as $(j,a)$ with $a = 1 \cdots 4$. The effective lattice is
shown in Fig.~\ref{scm2}$b$. Now the gauge constraint becomes $
\gamma_{j,1}\gamma_{j,2}\gamma_{j,3}\gamma_{j,4} (-1)^{n_j} = 1$.
Again we can map the Majorana fermions to $Z_2$ gauge field as
following:  \beqn \sigma^z_{j, j + x} =
i\gamma_{j,2}\gamma_{j+x,1}, \ \sigma^z_{j, j + y} =
i\gamma_{j,3}\gamma_{j+y,4}; \cr\cr \sigma^x_{j,j+y} = \prod_{k
\leq j} i\gamma_{k,4}\gamma_{k,3}, \ \ \sigma^x_{j,j+x} =
\prod_{k\leq j}i\gamma_{k,1}\gamma_{k,2}. \eeqn The entire
Hamiltonian can be parallelly generalized from its 1d counterpart
Eq. \ref{1dgauge} \beqn H = \sum_j U(n_j - \bar{n})^2 - \sum_{\nu
= x, y} t_2 \sigma^z_{j,j+\nu} \cos(\frac{\phi_j}{2} -
\frac{\phi_{j + \nu}}{2}). \label{2dgauge}\eeqn Again this
Hamiltonian is subject to the gauge constraint \beqn
\sigma^x_{j,j-x}\sigma^x_{j,j+x}\sigma^x_{j,j-y}\sigma^x_{j,j+y} =
(-1)^{n_j}.\eeqn

The 2D $Z_2$ gauge field is drastically different from 1D, in the
sense that it has a nontrivial liquid phase even when $\phi_j$ is
disordered. In the MI phase of SC islands, integrating out the
gapped fluctuation $\phi_j$ will induce gauge invariant dynamics
for $Z_2$ gauge field: \beqn H_{\mathrm{ring}} = \sum_j - K
\sigma^z_{j, j+x}\sigma^z_{j+x,
j+x+y}\sigma^z_{j+y,j+x+y}\sigma^z_{j,j+y}, \label{ring} \eeqn
with $K \sim t_2^4/U^3$. This is a standard ring exchange term of
$Z_2$ gauge field. This term favors the ring product of $\sigma^z$
around each unit plaquette to be 1. In ordinary $Z_2$ gauge field,
this ring exchange term will compete with the $Z_2$ string tension
term $\sum_{i,\nu} -h\sigma^x_{i, i+\nu}$, and when $K$ dominates
$h$ the system is in a $Z_2$ liquid phase with topological order
which cannot be described by local order parameter \cite{kitaev1}.
When $h$ dominates $K$, the system is in a confined phase without
topological order, where $Z_2$ charged matter is not just gapped,
but also confined spatially by a linear potential. Physically
these two phases can be understood by the behavior of ``vison",
which is a topological excitation with the product
$\prod_{\square}\sigma^z = -1$ on one plaquette. The vison carries
a global $Z_2$ charge, because one can only create/anihilate a
pair of vison by operator $\sigma^x_{i,i+\nu}$. In the liquid
phase, the vison number is conserved mod 2, while in the confined
phase the global $Z_2$ symmetry is spontaneously broken. This
effect is manifested in the dual description of the $Z_2$ gauge
field, which is formulated through the mapping: $\prod_{\square}
\sigma^z = \tau^x_{\bar{j}}$, $\sigma^x_{j,j+y} =
\tau^z_{\bar{j}-x}\tau^z_{\bar{j}}$. $\bar{j}$ denotes the dual
lattice shown in Fig.~\ref{scm2}$b$. Therefore the ordinary
quantum $Z_2$ gauge field is dual to a 2d transverse field quantum
Ising model $ H =\sum_{\bar{j}, \nu} -
h\tau^z_{\bar{j}}\tau^z_{\bar{j} + \nu} - K \tau^x_{\bar{j}}$.
When $K \gg h$, this Ising model is in the disordered phase, where
the $Z_2$ conservation of $\tau^x$ (vison number) is preserved;
when $h \gg K$, the $Z_2$ global symmetry of $\tau^z$ is
spontaneously broken.

\begin{figure}
\includegraphics[width=2.6in]{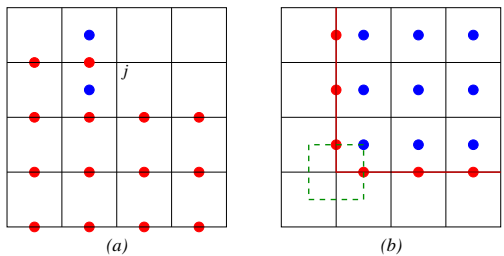}
\caption{$(a)$, the electron at site $j,1$ (Fig.~\ref{scm2}$a$)
can be represented as $\exp(i\phi_j/2)$ times a string of
$\sigma^z$, denoted by red circles; By binding two visons (blue
circles), the fermion becomes a boson. $(b)$, the Wilson loop
operator as a product of $\sigma^z$ (red circles), in the dual
formalism is a product of $\tau^x$ on the dual lattice (blue
circles). In the $Z_2$ liquid phase, the perturbation of $J$ term
will only change the Wilson loop at the corner, represented by the
green dashed square.} \label{scm3}
\end{figure}

In our case, operator $\sigma^x_{j,j+\nu}$ is a nonlocal product
of Majorana fermions, hence the string tension term
$\sigma^x_{j,j+\nu}$ cannot exist in the Hamiltonian. Therefore in
the disordered phase of $\phi_j$ (MI of SC islands), the local
vison number commutes with the Hamiltonian $i.e.$ vison is
completely static. Hence the $Z_2$ gauge field is in its liquid
phase. The $Z_2$ liquid phase is deconfined $i.e.$ an extra
electron will carry an infinite string of $\sigma^x$ due to the
gauge constraint, but the energy cost is finite. For instance, the
electron operator at site $j,1$ (Fig.~\ref{scm2}$a$) can be
written as \beqn c_{j,1} \sim e^{i\phi_j/2}\gamma_{j,1} \sim
\prod_{k \leq j} \sigma^z_{k - x, k }\sigma^x_{j - x,
j}e^{i\phi_j/2}, \label{electron2d}\eeqn where the product
includes all the $\sigma^z$ on $x-$links to the left or below site
$j$ (Fig.~\ref{scm3}$a$). $\sigma^x_{j, j+\nu}$ creates a pair of
vison excitations, and since in the $Z_2$ liquid state the vison
excitation is gapped and conserved, the least energy consuming
behavior of an injected electron is to form a bound state with two
visons, and become a charge-$e$ boson represented by following
operator: \beqn b_j \sim \prod_{k \leq j} \sigma^z_{k - x, k }
\exp(i\phi_j/2). \label{boson}\eeqn The SF phase can also be
viewed as a condensate of $b_j$. If an electron is injected into
this system, it will fractionalize into a mobile charge-$e$ boson
and a pair of static visons. If a Cooper pair is injected into
this system, it will fractionalize into two bosons, instead of two
electrons. In this liquid phase, the boson and the vison will have
mutual semion statistics $i.e.$ when a boson $b_j$ encircles a
vison through a close loop, the system wave-function acquires a
minus sign \cite{kitaev1}.

The SF phase has vortex with $hc/2e$ magnetic flux, which is bound
with a vison. If we start from the SF phase, the $Z_2$ liquid
phase can be viewed as a condensate of the double vortex with
$hc/e$ magnetic flux of SF phase, since the MI liquid phase still
has vison conservation, and the product of quantum circulation of
the vortex and the charge in the vortex condensate is a constant:
$Q_vQ_e = hc$ \cite{senthilfisher}. The $Z_2$ gauge field has no
gapless photon excitations, hence at the transition between SF
phase and the $Z_2$ liquid phase, the $Z_2$ gauge field does not
introduce further anomalous dimension to $\psi \sim \exp(i
\phi/2)$. For instance, when $\bar{n}$ is integer this transition
is a 3d XY transition described by order parameter $\psi$. 
However, the
physical Cooper pair operator $\psi^2 \sim \exp(i\phi)$ gains a
rather large anomalous dimension, which has been calculated by
various methods \cite{vicari2003}: $\eta_{\psi^2} \sim 1.47$.
Similar situations were discussed at the transition between $Z_2$
spin liquid and spiral order in frustrated magnets
\cite{senthil1994,xusachdev}.

There is another equivalent way of describing the fractionalized
$Z_2$ liquid phase $i.e.$ the disordered phase of $\phi$. Let us
assume the filling on every SC island is even, and we can define
Ising variables \beqn \mu^x_j = i\gamma_{j,1}\gamma_{j,3} = i
\gamma_{j,2}\gamma_{j,4}, \cr\cr \mu^y_{j} =
i\gamma_{j,3}\gamma_{j,2} = i \gamma_{j,1}\gamma_{j,4}, \cr\cr
\mu^z_{j} = i\gamma_{j,1}\gamma_{j,2} = i
\gamma_{j,4}\gamma_{j,3}. \label{dualmu}\eeqn $\mu^a$ satisfy the
algebra of Pauli matrices. Notice that unlike $\sigma^a$
introduced before, now the Ising variables $\mu^a$ are defined on
the sites of the lattice instead of the links. Now the ring
exchange term $H_{\mathrm{ring}}$ reads \beqn H_{\mathrm{ring}} =
\sum_{j} -K \mu^x_{j}\mu^y_{j+x}\mu^x_{j+y}\mu^y_{j+x+y}.
\label{wenform}\eeqn This is precisely the model introduce in
Ref.~\cite{wenmodel}, as a concrete example with topological
orders. This model is equivalent to the toric code model
\cite{kitaev1}, which becomes manifest once we switch the
definition of $\mu^x$ and $\mu^y$ in Eq.~\ref{dualmu} for one of
the two sublattices of the square lattice.


\section{two dimension with intra-island tunneling}

Now we consider a lattice structure slightly different from the
previous section, with the entire system compressed in the $y$
direction, until there is a considerable intra-island tunnelling
$- J i\gamma_{j,3}\gamma_{j,4}$. Using the mapping derived in last
section, this operator is $J\sigma^x_{j,j-y}\sigma^x_{j,j+y}$ in
terms of the $Z_2$ gauge variables, and this term becomes a ring
exchange on the dual lattice $ \sigma^x_{j,j-y}\sigma^x_{j,j+y} =
\tau^z_{\bar{j}}\tau^z_{\bar{j}-x}\tau^z_{\bar{j}-x-y}\tau^z_{\bar{j}-y}$,
which represents a pair hopping of visons. Notice that $\tau^z$
are defined on the sites of the dual lattice instead of the links.
The Hamiltonian of the dual Ising variables in the MI phase reads
\beqn H = \sum_{\bar{j}} - J
\tau^z_{\bar{j}}\tau^z_{\bar{j}-x}\tau^z_{\bar{j}-x-y}\tau^z_{\bar{j}-y}
- K \tau^x_{\bar{j}}. \label{xumoore}\eeqn This is precisely the
model studied in Ref.~\cite{xumoore} in the context of $p \pm ip$
superconductor Josephson array. The symmetry of this model is
quasi-local in the sense that we can reverse the sign of $\tau^z$
along any column or row arbitrarily, without changing the
Hamiltonian. Physically this symmetry means that the vison number
has $Z_2$ conservation along each row and column on the dual
lattice.

The phase diagram of model Eq. \ref{xumoore} has been studied both
analytically and numerically \cite{xumoore,vidal2009,gvidal2009}.
Because of its special self-duality structure, it is expected that
a transition occurs at precisely $J = K \sim t_2^4/U^3$
\cite{xumoore,xumoore2}. When $K > J$, the system is in the $Z_2$
liquid phase with topological order; when $J
> K$, the topological order vanishes and the ground state is nondegenerate,
although formally the nonlocal operator $\sigma^x$ has nonzero
expectation value. Due to the absence of string tension $ -
h\sigma^x_{j,j+\nu}$, $Z_2$ charged matter is deconfined in both
phases, which is very different from the ordinary $Z_2$ gauge
field. Mean field argument as well as numerical results suggest
that the transition at $J = K$ is first order
\cite{vidal2009,gvidal2009,hdchen2007}. In the $Z_2$ liquid phase,
the $J$ term enables the visons to move in pairs, therefore an
injected electron will fractionalize into two mobile parts:
charge-$e$ boson and vison pair. Notice that unlike the ordinary
$Z_2$ gauge field, this pair of visons cannot annihilate each
other, due to the quasi local conservation of vison numbers
discussed in last paragraph.

The difference between these two phases can be further
characterized by the Wilson loop. In the phase with $J > K$, since
the zeroth order ground state with $K = 0$ is an eigenstate of
$\tau^z$, the Wilson loop $\langle \prod_{\mathcal{C}}\sigma^z
\rangle = \prod_{\mathcal{A}}\tau^x$ can be calculated
perturbatively with expansion of $K$, and it falls off according
to an area law: $\langle \prod_{\mathcal{C}}\sigma^z \rangle \sim
(K/J)^{\mathcal{A}}$ \cite{fradkin1978}. Here $\mathcal{C}$ and
$\mathcal{A}$ represent a closed loop and the area enclosed inside
this loop respectively. In the $Z_2$ liquid phase with $K
> J$, it is usually expected that with the presence of transverse
field $\sum_j - h \sigma^x_{j,j+\nu}$ the Wilson loop falls off
with a perimeter law, which can also be revealed with a
perturbation of $J/K$ on the ground state with $J = 0$, where the
Wilson loop is a constant. However, In our situation with the
intra-island tunnelling, the first order expansion of $J$ term in
Eq. \ref{xumoore} will only change the Wilson loop at the corners
of loop $\mathcal{C}$ (Fig.~\ref{scm3}$b$). Therefore we expect
the Wilson loop in the $Z_2$ liquid phase falls off as a special
``corner law" $\langle \prod_{\mathcal{C}}\sigma^z \rangle \sim
e^{-\mathcal{N}J/K}$, $\mathcal{N}$ is the number of corners of
this Wilson loop $\mathcal{C}$.

Based on the analysis above, when $J$ is small, by reducing
$t_2/U$ from infinity we will first drive a transition from the SF
phase to a $Z_2$ liquid phase, and then enters an ``area law" MI
phase through a first order transition. When $J$ is large enough,
there can be a direct transition between the SF phase and the
``area law" phase. This transition can be viewed as proliferating
the $hc/2e$ vortices of the SF phase which can only move in pairs
due to the quasi-local conservation of visons. This type of paired
directional vortex dynamics was the key of the roton liquid phase
proposed before \cite{rotonliquid,bosemetal,xumoore2}, which is a
stable phase with gapless vison excitations and quasi one
dimensional dispersions. Therefore the direct transition in
Fig.~\ref{scm2}$d$ might expand into a stable roton liquid phase.
We will leave this possibility to future studies
\cite{xufufuture}.

If we turn on not only $- J^z_{j} i\gamma_{j,3}\gamma_{j,4}$, but
also $- J^x_{j} i\gamma_{j,2}\gamma_{j,3}$ and $- J^y_{j}
i\gamma_{j,3}\gamma_{j,1}$, after introducing Ising variables
$\mu^a$ as Eq.~\ref{dualmu}, the model describing the system
becomes \beqn H &=&  \sum_{j} -K
\mu^x_{j}\mu^y_{j+x}\mu^x_{j+y}\mu^y_{j+x+y} \cr\cr && + J^x_j
\mu^x_{j} + J^y_j \mu^y_{j} + J^z_j\mu^z_{j}. \label{model}\eeqn
It will be interesting to do a full analysis of all the possible
phases of this model with different choices of site dependent
transverse fields $J^a_{j}$. For instance, with $J^x = J^y = 0$,
and $J^z$ is site independent, this model reduces to the model in
Ref.~\cite{xumoore}. If $J^x_{A} = J^y_B \neq 0$ ($A$ and $B$ are
two different sublattices of the square lattice), while all the
other transverse fields are zero, this model is equivalent to the
toric code model with one component of transverse magnetic field,
and there is a confine-deconfine phase transition driven by this
transverse field.

\section{summary}

In summary, Josephson array with Majorana fermion zero modes
around each SC island is described by a precise $Z_2$ gauge field
and matter field formalism, which leads to unusual
fractionalization features in both one and two dimensions. Various
pseudo-spin models with topological phases can be realized with
the Josephson array. The fractionalization can be measured with
single electron tunnelling experiments, because an electron will
fractionalize into a boson and topological defects, and the single
electron green's function becomes a convolution of two fractional
excitations. This will be discussed in more details in another
paper. In our current paper we focus on the disordered phase of
the Josephson array with fractional excitations, but when $H_{t1}$
is nonzero or $\bar{n}$ is away from integer, many interesting
phase transitions can occur inside the SF phase, we will also
study these physics in future \cite{xufufuture}.

The authors are sponsored by the Society of Fellows, Harvard
University, and the Milton Funds.

\bibliography{junction1}

\end{document}